\documentclass[aps,prb,twocolumn,groupedaddress,amsmath,showpacs]{revtex4}
\usepackage{amsmath}
\usepackage{graphicx}
\begin{document}
\title{Layered ferromagnet-superconductor structures: the $\pi$ state
and proximity effects}
\author{Klaus Halterman}
\email{klaus.halterman@navy.mil}
\affiliation{Sensor and Signal Sciences Division, Naval Air warfare Center,
China Lake, California 93355}
\author{Oriol T. Valls}
\email {otvalls@umn.edu}
\affiliation{School of Physics and Astronomy and Minnesota Supercomputer
Institute, Minneapolis, Minnesota 55455}
\date{\today}

\begin{abstract}
We investigate clean mutilayered structures of the SFS and SFSFS type,
(where  the S layer is intrinsically superconducting and the F layer is ferromagnetic)
through numerical solution of the self-consistent Bogoliubov-de Gennes equations
for these systems. We obtain results for the pair amplitude, the 
local density
of states, and the local
magnetic moment. We find that as a function of the thickness $d_F$ of the 
magnetic layers separating adjacent superconductors,
the ground state energy
varies periodically between two stable states.
The first state is
an ordinary ``$0$-state", in which the order parameter has 
a phase difference of zero between consecutive
S layers, and the second is
a ``$\pi$-state", where the sign alternates,
corresponding to a phase difference of $\pi$ between adjacent S layers. 
This behavior
can be understood from simple arguments.  The density of states 
and the local magnetic moment reflect also
this periodicity. 

\end{abstract}
\maketitle

\section{Introduction\label{introduction}}
The study of layered ferromagnet-superconductor (F/S) heterostructure
has sustained
the active  interest of many researchers. This
is due in great part to continuing and recent progress in the preparation
and fabrication of multilayer systems, and to the
potential use  of such
heterostructures
in various important applications. In particular,  
structures consisting
of alternating ferromagnet (F) and superconductor (S) layers may
exhibit, in certain cases, a ground state
in which the difference $\Delta \phi$ between the order
parameter phase of adjacent superconductor layers equals $\pi$. 
These are the so called ``$\pi$ junctions''.
These F/S hybrid 
structures offer advances in the field of nanoscale technology, 
including
quantum computing,\cite{blatter}
where the implementation of a quantum two-level system is based on superconducting loops of 
$\pi$ junctions.
Furthermore, artificial composites involving a superconductor sandwiched between two 
ferromagnets, the design of which follows from  giant magnetoresistive (GMR) devices,
show potential use as spin-valves\cite{tagirov,mazin} and nonvolatile memory elements.\cite{oh}
An essential principle behind
many of these spin-based devices is the damped oscillatory nature of the 
Cooper pairs in the ferromagnet region, and the associated phase shift in the
superconducting order parameter.

The  coupling between nearby superconductors 
separated by a ferromagnet is a property that follows from
the proximity effects, which
in the context of F/S multilayers consist of the existence of 
superconducting correlations in the ferromagnet and magnetic
correlations in the superconductor,
arising from their mutual influence.
The resulting superconducting phase coherence is quantified by
the pair amplitude $F({\bf r})=\langle \hat{\psi}_{\downarrow}({\bf r})
\hat{\psi}_{\uparrow}({\bf r})\rangle$, where the $\hat{\psi}_\sigma$ are the
usual annihilation operators.
It is now well established that the leakage of superconductivity is due to the
process of Andreev reflection\cite{andreev}, whereby a quasiparticle incident on a F/S interface
is retroreflected as a quasihole of opposite spin. It 
is in turn the coherent superposition of these states, spin split by the
exchange field in the ferromagnet, that  ultimately leads to  damped oscillations
of $F({\bf r})$ in the magnet, with a characteristic length $\xi_F$ typically much smaller than 
the superconducting coherence length $\xi_0$.
These oscillations are akin to high field oscillatory
phenomena described a long  time ago.\cite{ff,lo}
In the absence of currents and magnetic fields,
the modulation of the order parameter  determines
whether two neighboring superconductor layers share
a stable $\pi$ or $0$ phase difference.
For  a multilayer F/S heterostructure with ferromagnet layers
of order $\pi \xi_F$ in width, it is intuitively evident,
taking into account
the continuity of $F({\bf r})$ across the F/S interface and
the particular oscillatory nature of the pair amplitude in the ferromagnet, 
that 
a configuration will result in which it  is energetically favorable to
have a phase difference of $\Delta \phi = \pi$, rather than zero, 
between successive 
superconducting layers. This indeed turns out to be the case.

Although recently there has been a surge
of interest in 
the study of F/S multilayer
structures, (see e.g. the theoretical work of
Refs.~\onlinecite{belzig,zyuzin,zenchuk,volkov,krawiec,radovic,buzdin2,buzdin3,demler,
bergeret,andreev2,buzdin4,bergeret2,krivoruchko,krivoruchko2,barash,radovic2,
triplet}
and the experimental work discussed below)
work on superconductor-ferromagnet-superconductor 
(SFS) Josephson junctions started long ago.\cite{parks} The Josephson current 
was calculated
for a short weak link in the clean limit,
and found to exhibit oscillations as a function of the ferromagnet exchange 
field.\cite{buzdin} It was later 
demonstrated that for an SFS sandwich obeying the dirty limit conditions, the 
critical current oscillates as a function 
of the thickness of the magnet, and of the exchange field.\cite{buzdin2}
A more detailed analysis of dirty $\pi$ junctions near the critical temperature 
allowed for differing transparencies of the ferromagnet-superconductor 
interfaces.\cite{buzdin3}
Many interesting phenomena have been proposed or
discussed. Calculations were 
more recently performed for an SFS junction with arbitrary impurity 
concentration. For non-homogeneous magnetization,\cite{bergeret,triplet}
the 
superconductor may exhibit a nonzero 
triplet component extending well into the magnet.
For quasi two-dimensional, tight-binding, 
F/S atomic-scale multilayers, the ground state 
was shown in some cases 
to be the $\pi$ state
\cite{andreev2},
and the density of states (DOS) 
exhibited prominent  features that depend critically on the exchange field 
and transfer integral parameters.\cite{buzdin4}
For multilayer structures consisting of two ferromagnets and an insulator 
sandwiched between two superconductors, an
enhancement of the Josephson current was predicted\cite{bergeret2,krivoruchko}
for antiparallel alignment of the magnetization in the ferromagnet layers.
Spin-orbit scattering\cite{demler,krivoruchko2} 
and changing the relative orientation angle of the
in-plane magnetizations\cite{barash} were
shown to significantly modify the behavior of the dc Josephson 
current. 

The  rapidly evolving  theoretical views
compounded with technological advancements which permit the
fabrication of well-characterized heterostructures,
has prompted a considerable
number of experimental investigations of $\pi$ coupling on 
several fronts. A study of the superconducting transition temperature for
F/S multilayers revealed oscillatory behavior as a function of ferromagnet
thickness.\cite{jiang} For $\pi$ junctions  involving
relatively weak ferromagnets, variations in temperature 
can induce a crossover from  $\Delta \phi=0$ to the $\pi$ state, and 
this was observed\cite{ryazanov}
as oscillations of the critical current versus temperature. The transition to the $\pi$ state
is also reflected in critical current measurements for Josephson junctions in 
which the
ferromagnet layer separating 
the two superconductors  was systematically varied.\cite{kontos} 
The superconducting phase was measured directly\cite{kontos2} using SQUID's made of $\pi$ 
junctions, demonstrating a half quantum flux shift in the diffraction pattern.
Direct evidence of 
the oscillatory behavior of the superconducting
correlations in the ferromagnet was found through
tunneling spectroscopy measurements which yielded 
inversions in DOS  for a thin ferromagnetic film,  
in contrast
with the behavior in a superconductor.\cite{kontos3} 

A common feature that pervades 
most of the theoretical work mentioned above
is the use of
quasiclassical formalisms, often compounded by
the neglect of self-consistency for
the space dependent pair potential,
$\Delta({\bf r})$.
These approximations do have the advantage of
providing an accessible and
efficient method to approximately calculate
properties of inhomogeneous superconducting systems,
while avoiding the cumbersome numerical issues that
arise when attempting to solve the corresponding, much more complicated,
self-consistent microscopic equations. 
The general underlying drawback of such approximations however,
is the elimination from consideration of 
phenomena at the atomic length scale given by the Fermi wavelength,
$\lambda_F$, as can be seen
in the derivation of  the Eilenberger equations.\cite{eilenberger}
Further approximations follow when the assumption
is made that the mean free path is
much shorter than $\xi_0$, in which case the Eilenberger equations
reduce to  the widely 
used Usadel equations.\cite{usadel}
The elimination of the relatively small length scales  poses problems 
for quasiclassical methods (even when self-consistent)
when interfacial scattering is involved,\cite{demler} or when the
 geometry or
potentials have sharp variations on the atomic 
scale. These issues, of increasing
experimental importance given the ever improving
quality of the experimental samples, often require nontrivial
effective boundary conditions that must supplement
the basic equations.
The problem worsens when
dealing with
multilayer structures, where
the successive reflections and transmission of
quasiparticles creates closed trajectories\cite{ozana} that
may render the quasiclassical approximation
scheme inapplicable.

In this paper we
investigate the proximity effect and associated electronic
properties of clean three-dimensional F/S multilayer structures 
comprised of alternating
superconductor and ferromagnet layers.
Our emphasis is on the study of the existence of pair potential
behavior of the $\pi$ type.
We implement a  complete self-consistent
microscopic theory that
treats all the characteristic length scales on an equal footing, and thus
can accommodate all quantum interference effects that are likely
to be pertinent. 
The problem will be solved from a wave function approach using the 
Bogoliubov-de Gennes (BdG) equations. To do so,
we extend an earlier method\cite{proximity}
used for a single F/S structure, to allow for a more complicated
geometry, consisting of an arbitrary number of
layers. Self-consistency is rigorously included, as
it has been demonstrated\cite{proximity,klaus}  that
this is essential in the study of the proximity effect at F/S interfaces.

We present in Sec.\ref{method} the geometry and the numerical approach
we take to solve the microscopic
BdG
equations and obtain the
self consistent energy spectra (eigenvalues and eigenfunctions). 
We also explain in some detail how the local DOS 
and the ground
state energy are calculated.
In Sec.\ref{results} we first present our results for SFS
structures and show examples
of the relevant quantities:  these include first of all
the pair amplitude, which
is used  to illustrate the cases in which either the $0$ or
$\pi$ state is energetically favored. The thickness of the F layer or layers turns
out to be the decisive parameter, with the zero and $\pi$
states periodically alternating in stability as this quantity varies.
The experimentally accessible DOS averaged
over a superconducting layer is next discussed:
results for both the sum and the difference of the up and down spin terms are
presented and their correlation with the zero or $\pi$
states demonstrated. Results for the local magnetic moment, which we
show how to calculate from the spin-dependent local DOS, are
also given: in the superconductor this quantity measures the penetration
of magnetic correlations. 
We analyze also, in a similar fashion, a more complicated five layer structure,
discuss the similarities and differences between the two geometries,
and the generalization of our results to
more complicated structures.
Finally in Sec.\ref{conclusions} we 
briefly summarize our results and discuss potential
experimental implications and future work.

\section{\label{method}Method}
\begin{figure}
\includegraphics[scale=.4,angle=0]{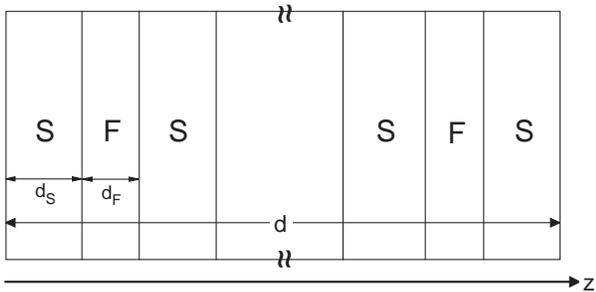}
\caption{\label{fig1} Schematic of the model geometry used in this paper.
The total thickness in the $z$
direction is $d$, and the thicknesses of the S and F layers
are $d_S$ and $d_F$ as indicated. There is a total number $N_L$
of layers $N_L=3$ and $N_L=5$ in
this work, with the outer ones being superconducting.
The line breaks in the middle region denote repetition.}
\end{figure}

In this paper
we consider a semi-infinite multilayer structure of total length $d$ in the
$z$-direction, consisting 
of an odd number $N_L$ of alternate superconductor (S) and ferromagnetic (F) 
layers, each of width $d_S$ and $d_F$
respectively (see Fig.~\ref{fig1}).  
The sandwich configuration is such that
the complete structure begins and ends with a superconductor layer. 
The free surfaces at $z=0$ and $z=d$ are specularly reflecting.
The basic methodology we use is an extension
of that which has been previously discussed.\cite{proximity,klaus}
Upon taking into account the translational invariance in the $x-y$ plane, one 
can immediately write down the  BdG\cite{bdg} equations
for the spin-up 
and spin-down quasiparticle and quasihole wave functions 
$(u_n^\uparrow,v_n^\downarrow)$,
\begin{equation}
\label{bogo}
\begin{bmatrix} {\cal H}-h_0(z)&\Delta(z) \\
\Delta(z) & -[{\cal H}+h_0(z)] \end{bmatrix}
\begin{bmatrix} u_n^{\uparrow}(z)\\v_n^{\downarrow}(z) \end{bmatrix}
=
\epsilon_n
\begin{bmatrix} u_n^{\uparrow}(z)\\v_n^{\downarrow}(z) \end{bmatrix}
\end{equation}
where the free-particle Hamiltonian is defined as,
\begin{equation}
\label{single}
{\cal H}\equiv -\frac{1}{2m}\frac{\partial^2}{\partial z^2} 
+\varepsilon_{\perp} 
-E_F(z).
\end{equation}
Here 
$\varepsilon_{\perp}$ is the transverse kinetic 
energy, the $\epsilon_n$ are the 
quasiparticle energy eigenvalues, and  $\Delta(z)$ is the pair potential, 
described below.
The magnetic exchange energy $h_0(z)$ is equal to a constant $h_0$
in the ferromagnet layers, and zero elsewhere.
A potential $U(z)$ describing interface
scattering can easily be added to Eqn.~\ref{single}.
We  define the quantity
$E_F(z)$ to
equal  $E_{FM}$ in the magnetic layers, 
so that in these regions, $E_{F\uparrow}=E_{FM}+h_0$, and 
$E_{F\downarrow}=E_{FM}-h_0$. Likewise, in the
superconducting layers,
$E_F(z)=E_{F S}$.
The dimensionless parameter $I\equiv h_0/E_{FM}$
characterizes the strength of the magnet. At $I=1$,
one therefore reaches the half metallic limit.
From the symmetry of the problem, 
the solutions for the other set of
wavefunctions  
$(u_{n}^{\downarrow},v_{n}^{\uparrow})$  
are  easily obtained  
from those of Eqns.~(\ref{bogo})
by allowing for both 
positive and negative energies.
The BdG equations are completed by the self
consistency condition for the pair potential,
\begin{equation}  
\label{del2} 
\Delta(z) =\frac{g(z)}{2} 
\sum_{\epsilon_n\leq \omega_D} \left[
u_n^\uparrow(z)v^\downarrow_n (z)+
u_n^\downarrow(z)v^\uparrow_n (z)\right]\tanh(\epsilon_n/2T), 
\end{equation} 
where $T$ is the temperature, 
$g(z)$ is the effective  coupling describing 
the electron-electron interaction, which be take to be a constant $g$ within
the superconductor layers and zero within the ferromagnet layers,
and $\omega_D$
is the Debye energy. We have not included spin-orbit
coupling, and assumed  that all of the $F$
layers are magnetically aligned and hence\cite{triplet} considered
singlet pairing only, in the $s$-wave.

We solve\cite{klaus} Eq.~(\ref{bogo})
by expanding the quasiparticle
amplitudes in terms
of a finite subset of a set of orthonormal basis vectors,
$u^{\uparrow}_n(z)=\sum_{q} u^{\uparrow}_{n q}\phi_q(z)$, and
$v^{\downarrow}_n(z)=\sum_{q} v^{\downarrow}_{n q}
\phi_q(z)$.
We use the complete set of eigenfunctions
$\phi_q(z)=\langle z \vert q \rangle = \sqrt{{2}/{d}}\sin(k_q z)$,
where 
$k_q = {q/\pi}{d}$,
and $q$ is a positive integer.
The finite range of the pairing interaction $\omega_D$
permits the number $N$ of such basis vectors
to be cut off in the usual way.\cite{proximity}
Once this is done,
we arrive at the following $2N\times2N$ matrix eigensystem,
\begin{equation}
\label{nset1}
\begin{bmatrix} H^+&D \\
D & H^- \end{bmatrix}
\Psi_n
=
\epsilon_n
\,\Psi_n,
\end{equation}
where 
$\Psi_n^T =
(u^{\uparrow}_{n1},\ldots,u^{\uparrow}_{nN},v^{\downarrow}_{n1},
\ldots,v^{\downarrow}_{nN}).$
The matrix elements  $H^+_{q q'}$ 
connecting $\phi_q$ to $\phi_{q'}$
are 
constructed from the real-space quantities in Eq.(\ref{bogo}),
\begin{widetext}
\begin{subequations}\label{basic}
\begin{align}
H^{+}_{q q'}&=\left\langle q \left\vert\Bigl[-\frac{1}{2m}\frac{\partial^2}
{\partial z^2} +
\varepsilon_{\perp}-E_F(z) \Bigr]
-h_0(z)\right\vert q' \right\rangle\, \nonumber \label{basic1}\\ 
&=\left[\frac{k^2_q}{2m} + \varepsilon_{\perp}\right]\delta_{q q'}
 -\int_0^{d} dz\, \phi_q(z) E_{F \uparrow} (z)\phi_{q'}(z) -
 \int_{0}^{d} dz\, \phi_q(z)E_{F} (z)\phi_{q'}(z).
\end{align}
The expression for $H^-_{q q'}$ is calculated similarly. 
The off-diagonal matrix elements $D_{q q'}$
are given as,
\begin{align}
D_{q q'}&=\left\langle \,q \left\vert \Delta(z) \right\vert q' \,\right\rangle  =\int_{0}^d dz \,\phi_q(z)\Delta(z)\phi_{q'}(z).
\end{align}
\end{subequations}
After performing the integrations, Eq.(\ref{basic1}) can be expressed
as
\begin{subequations}
\begin{align}
 H^{+}_{q q'}&=-\sum_{n=2}^{N_L-1}
\Biggl \lbrace \frac{E_{F\uparrow}}{d}\Bigl[ \frac{\sin[n(k_{q}-k_{q'})(d_F+d_S)/2]}{(k_{q}-k_{q'})}
- \frac{\sin[ (k_{q}-k_{q'})(n(d_F+d_S)/2-d_F)]}{(k_{q}-k_{q'})}  \nonumber \\
 &+ \frac{\sin[ (k_{q}+k_{q'})(n(d_F+d_S)/2-d_F)]}{(k_{q}+k_{q'})} 
-\frac{\sin[n(k_{q}+k_{q'})(d_F+d_S)/2]}{(k_{q}+k_{q'})} \Bigr]  \nonumber \\
&+ \frac{E_{FS}}{d}\Bigl[  \frac{\sin[ (k_{q}-k_{q'})(n(d_F+d_S)/2-d_F)]}{(k_{q}-k_{q'})} 
-\frac{\sin[n(k_{q}-k_{q'})(d_F+d_S)/2]}{(k_{q}-k_{q'})}
\nonumber \\
 &+ \frac{\sin[n(k_{q}+k_{q'})(d_F+d_S)/2]}{(k_{q}+k_{q'})} 
-\frac{\sin[ (k_{q}+k_{q'})(n(d_F+d_S)/2-d_F)]}{(k_{q}+k_{q'})} 
\Bigr] 
\Biggr \rbrace
, \qquad q\neq q',  
\end{align}
where the sum is over even integers only.
The diagonal matrix elements are somewhat simpler, and are written,
\begin{align}
H^{+}_{q q}&=\frac{k^2_q}{2m} + \varepsilon_{\perp}-
\frac{ E_{F\uparrow}}{2 d}\left[ {N_L d_F}-\frac{1}{ k_q} \sum_{n=2}^{N_L-1} 
\sin[n k_q (d_F+d_S)-d_F] -\sin[n k_q (d_F+d_S)] \right] \nonumber \\
&-\frac{ E_{F S}}{2 d}\left[ {N_L d_S}-\frac{1}{ k_q} \sum_{n=2}^{N_L-1} 
\sin[n k_q (d_F+d_S)] -\sin[n k_q (d_F+d_S)-d_F]\right].
\end{align}
\end{subequations}
The self-consistency condition, Eq.(\ref{del2}), is now transformed into,
\begin{equation}
\label{selfcon}
\Delta(z) = \frac{\pi \lambda(z)}{k_F d}
\sum_{p,p'}
{\sum_q}
\int{d\varepsilon_\perp}
\left[
u^\uparrow_{n p}v^\downarrow_{n p'}+
u^\downarrow_{n p}v^\uparrow_{n p'}\right]
\sin(k_p z)\sin(k_{p'} z)
\tanh(\epsilon_n/2T),
\end{equation}
\end{widetext}
where $\lambda(z)=g(z) N(0)$, and $N(0)$ is the DOS for both spins of the 
superconductor
in the normal state. The quantum numbers $n$  encompass
the continuous transverse energy $\varepsilon_{\perp}$,
and  the quantized
longitudinal momentum index $q$.

The primary quantity of interest is the local density of
one particle excitations in the system, $N(z,\varepsilon)$. 
Current experimental tools such
as the scanning tunneling microscope (STM) have atomic
scale resolution, and make this quantity experimentally accessible.
Since it is assumed that  well defined
quasiparticles exist, the tunneling current is simply expressed 
as a convolution of the one-particle spectral function of the STM tip
with the spectral function for the ferromagnet-superconductor system.\cite{gygi}
The resultant tunneling conductance, which
is proportional to the DOS, is then
given as a sum of the individual contributions to the DOS 
from each spin channel. We have:
\begin{equation}
\label{tdos}
N(z,\varepsilon)= N_\uparrow(z,\varepsilon)+N_\downarrow(z,\varepsilon), 
\end{equation}
where
the local DOS for each  
spin state is given by
\begin{subequations}\label{dos}
\begin{align}
{N}_\uparrow(z,\epsilon) 
&=-\sum_{n}
\Bigl\lbrace[u^\uparrow_n(z)]^2
 f'(\epsilon-\epsilon_n) 
+[v^\uparrow_n(z)]^2
 f'(\epsilon+\epsilon_n)\Bigr\rbrace, \\ 
{N}_\downarrow(z,\epsilon) 
&=-\sum_{n}
\Bigl\lbrace[u^\downarrow_n(z)]^2
 f'(\epsilon-\epsilon_n) 
+[v^\downarrow_n(z)]^2
 f'(\epsilon+\epsilon_n)\Bigr\rbrace.
\end{align}
\end{subequations}
Here thermal broadening is accounted for
in the term involving the derivative of the
Fermi function $f$, $f'(\epsilon) = \partial f/\partial \epsilon$.

We shall see below that we will also need to
compare different self-consistent
states. In general this is done
in terms of  the free energy. 
However, we will consider here only the low temperature limit.
For $T\rightarrow 0$, the entropy
term can be neglected, as it vanishes proportionally to $T^2$.
In this case all we need is the ground state energy $E_0$. 
In evaluating
this quantity some care must be taken in properly
including all energy shifts, even in the bulk
case\cite{tinkham,fw}. In the inhomogeneous
case the result\cite{kos} can be written as:
\begin{equation}
E_0 = \int^d_0 dz 
\int^0_{-\infty}  \epsilon N(z,\varepsilon) d\epsilon 
+ \frac{1}{g}  \langle|\Delta(z)|^2\rangle
\end{equation}
where the angular brackets in $\langle|\Delta(z)|^2\rangle$ denote the spatial average,
and $N(z,\varepsilon)$ is given in Eqn.~\ref{tdos}.
One can rewrite $E_0$ in a somewhat more standard way:\begin{equation}
E_0=-\sum_{p}{\sum_{n}}' \epsilon_n \Bigl[
(v^\uparrow_{n p} )^2+(v^\downarrow_{n p} )^2\Bigr] 
+ \frac{1}{g}  \langle|\Delta(z)|^2\rangle,
\label{final}
\end{equation}
which in principle gives $E_0$ in terms of the calculated excitation spectra.

\section{\label{results}Results}

In this section we present and discuss the results that we have obtained through
our numerical solution of 
the matrix eigensystem Eq.~(\ref{nset1}) and 
the self-consistency condition Eq.~(\ref{selfcon}).
We will study the two cases of $N_L=3$ and $N_L=5$, that is,
SFS and SFSFS structures separately. We consider
only ``regular'' structures in which all S layers have
the same thickness $d_S$, which we will take to
be a fixed value larger than $\xi_0$, while the F layers, when there is
more than one, have all the same  thickness, $d_F$, which we will  vary.
With the assumption that no current 
flows across the sample, the quantity $\Delta(z)$ can be taken to
be real, but it can in principle switch sign (zero or $\pi$ state) in going from
one S layer to the next. 

In our calculations
we have studied two different values of
the parameter $I$, $I=0.5$  and $I=1$. 
We have set the superconducting correlation length $\xi_0$
to $\Xi_0 \equiv k_{S}\xi_0=50$, where $k_S$ is the Fermi
wavevector of the superconductor, and taken 
$\omega \equiv \omega_D/E_{FS}=0.1$ for the dimensionless Debye energy cutoff.
It follows from  previous studies\cite{klaus} that the first 
of these parameters 
simply sets the overall length scale
in the superconductor and is of little relevance
whenever $d_S$ exceeds $\xi_0$, as will be the case here,
while the second is
unimportant at low temperatures (the limit that we will
consider), as it simply sets
the scale for $T_c$. We have also assumed that there is no 
oxide barrier between
the layers and that the ``mismatch parameter'' $\Lambda=(E_{FM}/E_{FS})$
is unity. A nonzero barrier height would in general diminish the amplitude
of all the phenomena discussed here, without qualitatively altering 
the results. The possible influence of varying $\Lambda$ is more complicated:
this parameter\cite{klaus} determines, together with $I$, the basic
spatial periodicity  of the problem, $\xi_F \approx (k_\uparrow-k_\downarrow)^{-1}$,
(where $k_\uparrow$ and $k_\downarrow$ are respectively the Fermi wavevectors
of the parabolic up and down spin bands in the ferromagnet)
which we shall
see is very important here. Furthermore, the amplitude of the oscillatory
behavior found in simpler SF structures decreases\cite{klaus} with $\Lambda$. 

As explained in previous work (see Refs.~\onlinecite{proximity,klaus}), 
the self consistent solution
to these equations is obtained  iteratively: one makes a suitable initial
guess for $\Delta(z)$, diagonalizes the system Eqn.~(\ref{nset1}) for
that guess, and computes an iterated $\Delta(z)$ from (\ref{selfcon}). The
process is then repeated until convergence is obtained.
The technicalities for the self consistent solution of these equations
were extensively discussed in previous work\cite{klaus,thesis}. 
The diagonalization in terms of the orthonormal basis chosen
must be performed for each value of $\varepsilon_\perp$ in the appropriate
range.
We took here $N_\perp=5000$ different
values of $\varepsilon_\perp$, except
as indicated below, and the number
of basis functions 
required for convergence was up to $N=1000$. The self consistent solution
process is terminated when the relative error between consecutive iterated
values of $\Delta(z)$ nowhere exceeds $10^{-4}$. We have found that 
the number of iterations needed to achieve self consistency can be quite large:
in most cases,
it exceeds fifty.

Because our objective here is to
discuss the possible $\pi$ states,
in starting the iteration process
we make two different initial guesses: one is of the ordinary ``zero'' state
form, where the initial guess has the same sign (conventionally
positive) in all the superconducting 
layers, and one of the $\pi$ form, where it alternates sign
from one S layer to the next.  We
have found that in some cases, for example for SFS 
structures with small $d_S$ (i.e., $\lesssim\xi_0$),
and $d_F \lesssim \pi \xi_F$, the
self-consistent $\Delta(z)$ typically converges to {\it either} a $0$
{\it or} $\pi$-state regardless of the initial guess, depending
on $d_F$. 
A similar trend holds for the small $d_S$ five layer SFSFS system
but over a  broader
$d_F$ range.
However, for the regular structures that we will focus on here,
with $d_S \gg \xi_0$, two different self consistent
solutions are { \it always} obtained, one of the zero
and one of the $\pi$ type,
according to the type of initial guess. We interpret this
as showing that two local minima of the free energy exist. We then 
have to determine
the stable minimum by calculating the free energy (or rather,
at low temperature, the ground state energy) of both self consistent
states, as discussed below, and comparing them.

\subsection{\label{SFS}SFS}

We consider first the case of an SFS sandwich. Preliminary investigations
showed that the situation of interest occurs  when the magnetic layer is not
too thick. This is as expected, since the overall length over which 
the superconducting
correlations penetrate (in an oscillatory way) into the magnet is 
characterized by $(k_\uparrow-k_\downarrow)^{-1}$, which
at the relatively large values of $I$
considered here is fairly small. Thus, we have taken in the studies presented
here a thickness $k_{S}d_S=300$ for the superconducting layers, and
the parameter
$k_{S} d_F$ meanwhile, is varied in the range between one and twenty.
The choice of $k_{S}d_S$  determines, through standard BCS theory relations,
the value of the ratio of the superconductor Fermi energy  to the bulk
order parameter.

\begin{figure}
\includegraphics[scale=.5,angle=0]{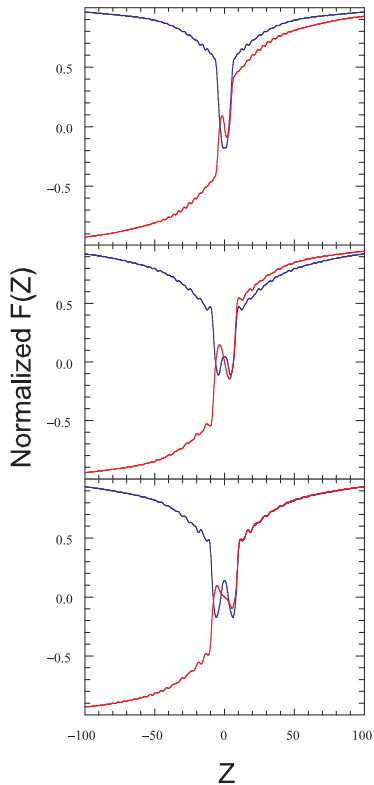}
\caption{\label{fig2} (Color online). 
Results for the pair amplitude $F(Z)$, normalized to the bulk superconductor
value, as a function of $Z\equiv k_{S}z$, in an SFS structure. The 
dimensionless thickness
of the S portions is $k_S d_S=300$, while the corresponding
values of the dimensionless thickness of the intervening F layer
are (from top to bottom) $k_S d_F=10, 16, 19$. The blue (solid) lines
represent self-consistent solutions of the zero type, and the red
lines alternative
self consistent solutions of the $\pi$ type. The value of $I$ is 0.5 and the
superconducting correlation length is $\xi_0=50$.}
\end{figure} 

As explained above, results were obtained by iteration
from  two initial configurations
of $\Delta(z)$, with the initial guesses corresponding
to opposite signs for the pair potential  in each
of the two S layers. 
For  the range of parameters considered here, 
both initial guesses led in all
cases to self consistent configurations, which were
either of the zero or of the $\pi$ types, according to the initial guess.
This is described in Fig.~\ref{fig2},
where we show 
examples of the two
self consistent solutions
for the pair amplitude $F(Z)$, as a function of the dimensionless
distance $Z\equiv k_S z$. 
It is not surprising that both types
of solutions are found: it is after all obvious that
in the limit where $d_F$ is sufficiently large, both
solutions must exist and be degenerate.
The pair amplitude $F(z)=g\Delta(z)$ does not
vanish identically in the magnetic region, but it exhibits the well-known 
oscillations. In the superconductor, it rises in absolute value
towards the bulk result, away from the S/F interfaces.
Results are shown for two values of $I$ and two
values of $k_S d_F$. We can see that in certain cases, depending
on the thickness of the F layer, $F(z)$ in the superconductor
(and hence $\Delta(z)$) is larger, in absolute value, for the 'zero' than for 
the $\pi$ state (see top panel), while in some other cases 
(middle panel) the opposite occurs, and for yet some other $d_F$ values
(see the bottom panel) there is no observable difference. Intuitively,
this happens because of the different way, depending on 
$k_S d_F$, in which the pair amplitude in the two superconductor regions must
adjust itself to the oscillations in the magnet. That the 
oscillatory behavior of the pair amplitude
in the F layer is clearly
different for the zero and $\pi$ solutions can be seen
by careful examination of the portion of the plots which lies in the
F region.

To
find out the most stable configuration, one
must compute the difference in the
ground state energies, or equivalently the
condensation energies, of the zero and $\pi$ states.  This can
be done in principle by using Eqn. (\ref{final}). In practice this
is computationally very difficult: the value of $E_0$ for each state must
be computed separately from its own spectrum (which
can consist of up to $10^6$ eigenstates), and the results subtracted. Since 
each $E_0$ includes the normal state energy, which is many orders of magnitude
larger than the condensation energy  sought, 
this requires extreme
numerical accuracy. The problem is exacerbated because the ``logarithmic''
last term in the right side
of Eqn. (\ref{final}) is in itself much larger
than the condensation energy (and the latter is itself considerably larger,
as we shall see below, than the condensation energy
{\it difference} between the two states),
and must be exactly canceled by a portion
of the first term. This is a well-known problem even in the bulk case,
where great care has to be taken\cite{tinkham} to make
the delicate cancellation analytically explicit. We have
found it technically
impractical to numerically compute $E_0$ from Eqn. (\ref{final})
for all cases considered\cite{altm} with
the required precision.
However, by using  increased values
of $N_\perp$ and $N$ 
in a few selected cases we have been able to verify that the ground
state {\it condensation} energy (that is, after subtracting the normal 
ground state energy $E_{0n}$ calculated for the same geometry and
parameter values except for setting $g=0$)
for either the zero or $\pi$ states is, for the cases
considered here where $d_F$ is small and $d_S>>\xi_0$,
approximately given by:
\begin{equation}
E_0-E_{0n}\approx - \alpha N(0) \langle|\Delta|^2\rangle. 
\label{app}
\end{equation}
This result is, {\it a posteriori} not surprising at all in the
limit of large $d_S$ and small $d_F$, as it is
quite similar to what is
found\cite{tinkhamsp} analytically for the bulk:
in that case $\alpha$ is exactly 0.5 and the spatial average
is of course replaced by the uniform bulk value. In our case
we find the coefficient $\alpha \lesssim 0.5$ within our numerical
uncertainty. The right
side of Eqn.~(\ref{app}) is of course very easy to compute. Thus,
we have adopted a procedure based on Eqn.~(\ref{app})
to compare condensation energies
for the two competing states.

\begin{figure}
\includegraphics[scale=.4,angle=0]{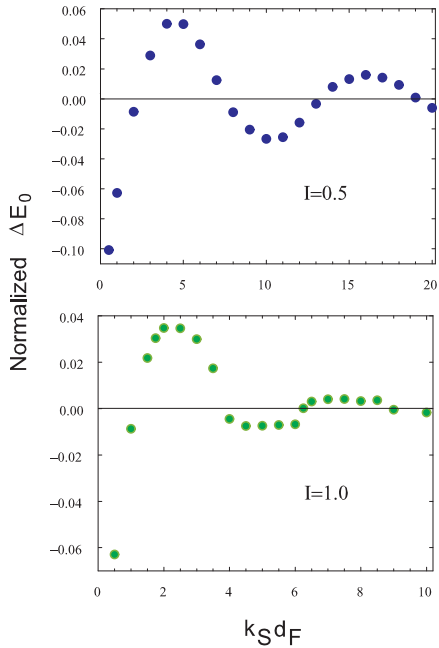}
\caption{\label{fig3} (Color online). The  
difference in condensation 
energies, $\Delta E_0$,
between the zero and $\pi$ states for an SFS
sandwich in the low temperature  limit,
normalized
to $N(0)\Delta_0^2$, calculated as explained in the text.
The results are 
plotted as a function of the dimensionless thickness $k_S d_F$ of the
ferromagnetic layer,  for two values of $I$. At small
$d_F$ the zero state is favored. The
periodicity of the results is determined by $(k_\uparrow-k_\downarrow)^{-1}$,
as expected. }
\end{figure} 

The results are shown in Figure \ref{fig3}.
The quantity plotted there is the 
difference
between the values of $\langle|\Delta|^2\rangle$ 
for the zero and $\pi$ states 
normalized to $N(0)\Delta_0^2$, where $\Delta_0$
is the bulk gap. This normalization corresponds to twice the bulk value
limit of the condensation energy. 
This is then a dimensionless measure of the
condensation (or equivalently, ground state) energy difference 
between the  self consistent  zero and $\pi$
configurations,  
(see Eqn.~(\ref{app})).
This normalized energy difference 
is plotted as a function of the 
dimensionless thickness $k_S d_F$ of the intermediate
F layer, which is sandwiched between thick ($k_S d_S=300$)
S layers.  We see that the difference
in energies is, as one would expect, only a small fraction (about
one tenth at the most) of the bulk
condensation energy. We also see that it is an oscillatory function 
of $k_S d_F$.
Comparison of the top and bottom panels (which correspond to $I=0.5$
and $I=1$ respectively) shows that the rough
periodicity of these results is approximately given
by $(k_\uparrow-k_\downarrow)^{-1}$, and it is in fact very similar\cite{klaus} 
quantitatively to the oscillatory
behavior of the pair amplitude $F(z)$ in a thick magnetic layer. At
small $k_S d_F$, the zero state is obviously very favored, as one
would expect, while in the limit
of large $k_S d_F$ the energy difference is of course  zero,
reflecting the degeneracy of the two states. The influence
of the parameter $I$ is quite dramatic: in the half metallic case
(lower panel) the first peak favoring the $\pi$ state is more prominent
and the zero state is generally speaking less favorable than for the
intermediate value of $I$ shown in the top panel.

\begin{figure}
\includegraphics[scale=.5,angle=0]{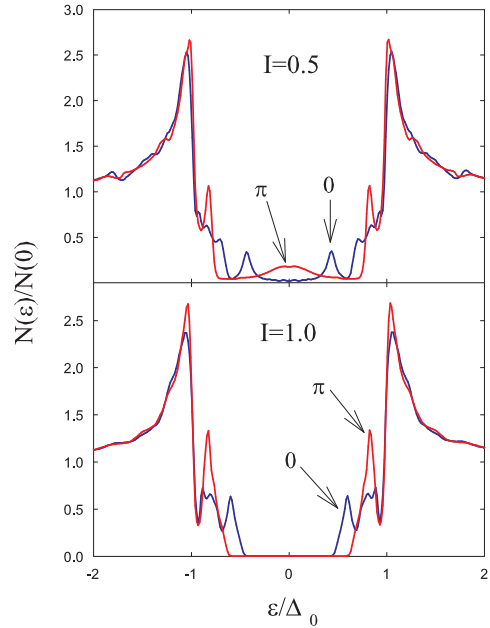}
\caption{\label{fig4} (Color online). Density of states (DOS) results for SFS
structures. The quantity plotted is the local DOS integrated over
one S layer, normalized to $N(0)$. The energy is normalized to the
bulk gap $\Delta_0$. The top panel shows results at $k_S d_F=5$, where the
stable state (see Fig.~\ref{fig3})
is of the $\pi$ type (red  curve, labeled as $\pi$) and at
$k_S d_F=10$, where the zero state is more stable (blue solid 
curve, labeled ``0'').
In the bottom panel, $I=1$ and consistent
with the doubling of $I$, the thicknesses 
displayed are halved to $k_S d_F=2.5$ ($\pi$
case) and $k_S d_F=5$ (zero case). See text for discussion.}
\end{figure} 

We turn now to the density of states (DOS) for this geometry.
Typical results are exhibited in Fig.~\ref{fig4}, where we show the DOS,
integrated over the superconducting region of thickness $k_S d_S=300$, as
a function of the energy normalized to $\Delta_0$. 
Results are shown for two values of $I$ (top and
bottom panels) and, for each value of $I$, at two values of
$k_S d_F$, one corresponding to the case where the  equilibrium state is
of the zero type, and the other corresponding to the opposite
situation. One can see that for $I=0.5$ there are states in the gap, and
that these states are more prominent in the $\pi$ case where
there is a zero energy small peak. At $I=1$ the DOS
results are also different: although for both of the cases shown there
is a gap in the spectrum, the location of the peaks near the gap edge
is not the same for the zero and $\pi$ states, with the first peaks being
more prominent and at higher energies in the latter case. Thus, there are
genuine differences between the DOS of zero and $\pi$ states, which may
be experimentally observable.

\begin{figure}
\includegraphics[scale=.5,angle=0]{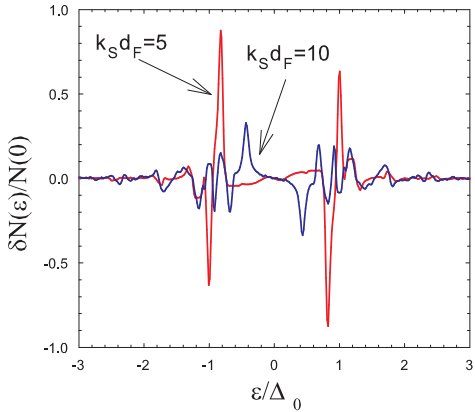}
\caption{\label{fig5} (Color online). Differential
density of states (DOS) between up and down spin states
for  SFS
structures. The quantity plotted is 
that defined in Eq.~(\ref{diffdos}), integrated over
one S layer, and normalized to $N(0)$. 
Results are shown for $I=0.5$ at $k_S d_F=5$, where the
stable state (see Fig.~\ref{fig3})
is of the $\pi$ type (red  curve) and at
$k_S d_F=10$, where the zero state is more stable (blue 
solid curve).
}
\end{figure}

It is also of interest to show the difference between the local DOS for up and
down states, as defined by 
\begin{equation}
\label{diffdos}
\delta N(z,\varepsilon) \equiv N_\uparrow(z,\varepsilon)-N_\downarrow(z,\varepsilon).
\end{equation}
This is done
in Fig.~\ref{fig5}, where results are shown for the two cases corresponding
to those also displayed in the top panel of Fig.~\ref{fig4}. The
differential DOS shown is integrated over the thickness of one S layer, and
normalized to the total normal bulk DOS value. Because of the
finite value of $I$, the results are not symmetric around zero energy.
One can see that the energy structure 
at the gap edge is appreciably more prominent for
the thickness value which corresponds to an equilibrium $\pi$ state,
while for the  zero state the structure is broader and more diffused.

An alternative way of illustrating the magnetic polarization effects, which
has also the advantage of providing local information, is through the use
of the local magnetic moment $m(z)$. This quantity is easily obtained by 
integration of the local DOS results.  One has:
\begin{equation}
\label{mm} 
m(z)= \mu_B\int d\varepsilon \delta N(z,\varepsilon)  f(\epsilon),
\end{equation}
where $\mu_B$ is the Bohr magneton and the integral extends over the occupied
states in the band. 
This can be cast in a more convenient form
as:
\begin{equation}
\label{mm2}
m(z)=\mu_B\left[\langle n_{\uparrow}(z) \rangle - \langle n_{\downarrow}(z) \rangle\right],
\end{equation}
where $\langle n_{\sigma}(z) \rangle$ is
the average number density for each spin subband, and is written
in terms of the quasiparticle amplitudes as,
\begin{equation}
\langle n_{\sigma}(z) \rangle= \sum_{n}
\Bigl\lbrace[u^\sigma_n(z)]^2
 f(\epsilon_n) 
+[v^\sigma_n(z)]^2[1-f(\epsilon_n)]\Bigr\rbrace,\qquad \sigma=\uparrow,\downarrow.
\end{equation}
It is more instructive to plot $m(z)$ normalized to
the corresponding integral of 
$N_\uparrow(z,\varepsilon)+N_\downarrow(z,\varepsilon)$. We denote this
normalized quantity by $M(z)$ and we plot it, in units of the Bohr magneton,
\begin{figure}
\includegraphics[scale=.5,angle=0]{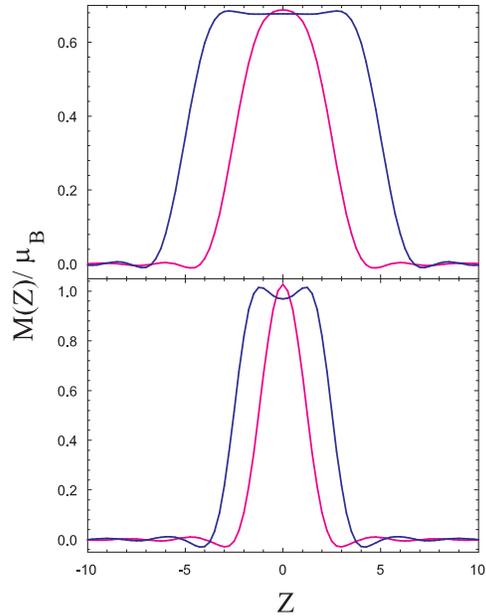}
\caption{\label{fig6} (Color online). Normalized local magnetic
moment as defined in the text and Eqn.~\ref{mm2}.  
Results in the top and bottom panels correspond to the same values of
$I$ and thickness as in the corresponding panels of Fig.~\ref{fig4}. Thus
the top panel is for $I=0.5$ and $k_S d_F=5, 10$, while the bottom panel
is for $I=1$ and $k_S d_F= 2.5, 5$. 
}
\end{figure}
in Fig.~\ref{fig6}. The two panels there correspond to
values of $I$ and $k_S d_F$ as in the corresponding
top and bottom panels of Fig.~\ref{fig4}.
We see in this figure that for relatively small $k_S d_F$, the quantity plotted
rises up sharply from the F/S interface and then has a slow modulation
as it approaches its bulk value in the F layer.
The magnetization  does not vanish identically
inside the superconductor: its behavior there consists of strongly
damped oscillations, with an overall characteristic spatial decay on the
order of a few Fermi wavelengths. The effect does not seem to depend
strongly on whether one is dealing with zero or $\pi$ states.

The self consistent results displayed can also be interpreted
as representing an effective, local value of $I(z)$, through the relation
$M(z)=\mu_B[(1+I(z))^{3/2}-(1-I(z))^{3/2}]/[(1+I(z))^{3/2}+(1-I(z))^{3/2}]$.
The quantity $I(z)$ is then the magnetic counterpart of the self consistent 
$F(z)$, measuring directly the magnetic part of the proximity effect, that is,
the leakage of magnetic correlations into the superconductor.

\subsection{\label{SFSFS}SFSFS}

In this subsection we consider the case of more complicated, five layer 
structures. These are realizable experimentally\cite{jiang}
and therefore of considerable interest. As in the
three layer case, we will study the situation where the 
three superconducting layers are relatively thick, taking again $k_S d_S=300$
and  the F
layers are thin enough so that F/S proximity effects cannot be neglected.

\begin{figure}
\includegraphics[scale=.5,angle=0]{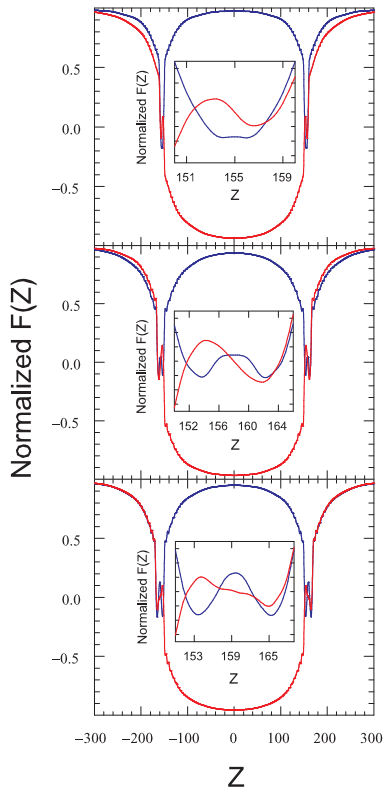}
\caption{\label{fig7} (Color online). Results for the pair amplitude 
$F(Z)$, normalized to the bulk superconductor
value, as a function of $Z\equiv k_{S}z$, in an SFSFS structure,
for $I=0.5$. The 
dimensionless thickness
of the S portions is $k_S d_S = 300$, and the corresponding
values of the dimensionless thickness of the intervening F layer
are (from top to bottom) $k_S d_F=10, 16, 19$. The blue (solid) lines
represent self-consistent solutions of the zero type, and the red
 lines of the $\pi$ type. The insets are a magnification of
one of the F regions. The  vertical axis in the insets varies
between $\pm .35$ in  dimensionless units.
}
\end{figure} 

We begin by considering (see Fig.~\ref{fig7}) the pair amplitude
$F(z)$. This figure
is in every way analogous to Fig.~\ref{fig2}, except for the insets, where
we display in more 
detail the behavior of $F(z)$ in one of the
ferromagnetic layers. Results for  solutions of both the
zero and the $\pi$ type are shown.  Both are obtained self-consistently,
the first by starting from an initial guess in which the sign of the
order parameter in the three S layers is always the same, and the second
by starting with a guess in which the order parameter in the middle
S layer is the opposite to that in the other two layers. Self consistent
solutions are always reached by iteration, for large $d_S$ and $d_F$
in the ranges shown, in either case.
We observe the expected depletion of $F(z)$ near the F/S
interfaces, and the subsequent approach towards it bulk
value over the
length scale $\xi_0$, with the maximum
$\Delta(z)$ in the central S layer slightly reduced from the bulk $\Delta_0$.
We also see in the main panels that depending on the
value of $k_S d_F$, the absolute value of $F(z)$ in the
superconductors varies periodically
between being larger in the zero state to being larger in the $\pi$
state, as was the case for three layer structures.  The insets 
illustrate more clearly how the existence of the two states relates
to the oscillations of $F(z)$ in the ferromagnetic region,
which are very different in each case.

\begin{figure}
\includegraphics[scale=.4,angle=0]{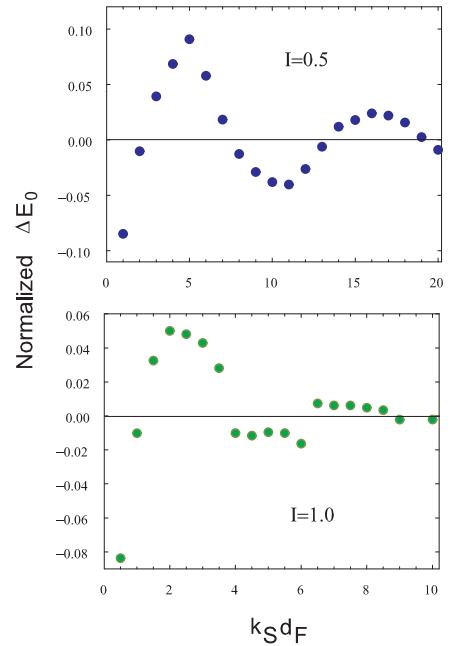}
\caption{\label{fig8} (Color online). Difference in condensation energies, $\Delta E_0$,
between
the zero and $\pi$ states for a five layer SFSFS  system, 
calculated as explained 
in the text, and normalized to $N(0)\Delta_0^2$, as in Fig.~\ref{fig3}.
This quantity is plotted as a function of the thickness $k_S d_F$ of each
ferromagnetic layer. Results for two values of $I$ are shown. }
\end{figure}

As in the three layer case, therefore, we find that there are two local
minima of the free energy, corresponding to the zero and $\pi$ 
alternatives. Again,  the
absolute minimum, at low temperature, must be found by comparing the two
condensation energies. This we do in the same way as for the three
layer case (see Eqn.~\ref{app} and associated discussion). The results are
shown in Fig.~\ref{fig8}, which should be compared with Fig.~\ref{fig3}.
The two figures are remarkably similar. In both cases the behavior
is oscillatory, with the same approximate spatial periodicity related
to that of the pair 
amplitude oscillations.
Again, the obvious results that the zero state is favored at small
$k_S d_F$ and that the two states are degenerate 
for large $k_S d_F$ are recovered.
The three and five layer plots are not identical, however: in the
latter case we find that the overall
scale of the phenomenon is nearly a factor
of two higher, as one can see by comparing
the vertical axes. The
first
peak favoring the $\pi$ state is higher and sharper for five
layers. Although the
effect of increasing the layer number is not as dramatic as 
that of increasing $I$, one can nevertheless assert from the trend that
the oscillatory behavior with $d_F$ would not only persist but would be
even more prominent if the number of layers
were further increased, as in superlattices.

\begin{figure}
\includegraphics[scale=.5,angle=0]{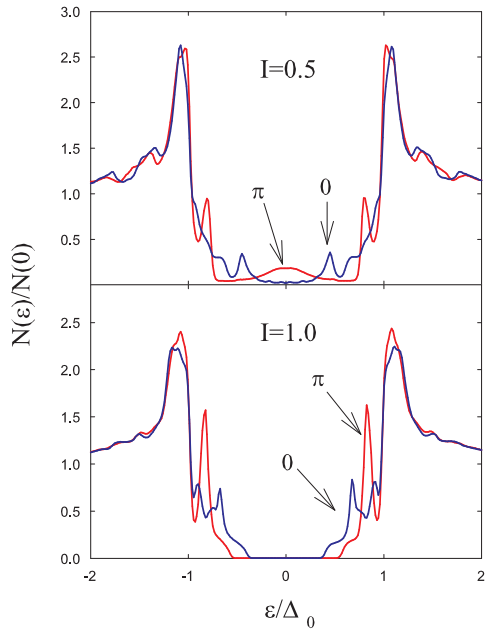}
\caption{\label{fig9} (Color online). Density of states (DOS) results for SFSFS
structures. The local DOS integrated over
one of the external S layers, normalized to $N(0)$, is plotted
vs. the energy normalized to the
bulk gap $\Delta_0$. The top panel shows results at $k_S d_F=5$, where the
stable state (see Fig.~\ref{fig8})
is of the $\pi$ type (red curve, labeled as ``$\pi$'') and at
$k_S d_F=10$, where the zero state is more stable (blue solid 
curve, labeled ``0'').
In the bottom panel, $I=1$ and 
the thicknesses are  $k_S d_F=2.5$ ($\pi$
case) and $k_S d_F=5$ (zero case). }
\end{figure} 

A few selected DOS results for
this SFSFS geometry are shown in Fig.~\ref{fig9}, which should be viewed in
comparison with the analogous Fig.~\ref{fig4} for the SFS structure. The
quantity plotted is averaged over one of the two outside S layers, and
all parameters are chosen to be the same as in Fig.~\ref{fig4}. The
similarity between the two figures is at first sight very remarkable, although
a second look shows that the structure of the subgap peaks is far from
being the same, particularly for the zero state case, where
additional shoulders appear at $I=1$. One concludes again
that many features, including the zero energy peak in the stable $\pi$ 
state of the top panel, are robust with respect to increasing the number
of layers, and very likely to persist, and even 
be more obvious, in larger regular structures.
 
\begin{figure}
\includegraphics[scale=.5,angle=0]{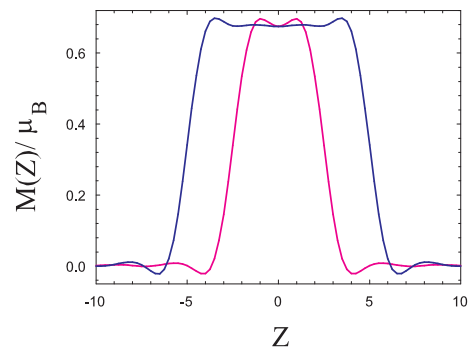}
\caption{\label{fig10} (Color online). Normalized local magnetic
moment for an SFSFS structure, as defined in the text and Eqn.~\ref{mm}.  
Results are  for $I=0.5$ and $k_S d_F=5, 10$.
}
\end{figure}

The differential DOS between up and down
states for this geometry exhibits a behavior sufficiently
similar to that displayed in Fig.~\ref{fig5} for the SFS case that there
is no need to display it in a separate figure here. On the other hand, it is 
worthwhile to illustrate an example of the 
normalized local magnetic moment $M(z)$.
This is done on Fig.~\ref{fig10}, where this quantity, as defined
in Eqn.~\ref{mm} is plotted with the same normalization and parameter values
as in the top panel of Fig.~\ref{fig6}. The behavior for the
two geometries is certainly
similar, but one again sees that the magnetic penetration effects become
more prominent as the number of layers increases from three to five. This
is another indication that such effects are very likely to be easier
to observe in  structures  involving a larger number of layers.

\section{conclusions \label{conclusions}}
We have rigorously investigated the proximity effects that
occur in 
clean multilayered F/S structures of the SFS and SFSFS
type. We used a microscopic wavefunction approach that does not coarse grain over 
length scales of order $\lambda_F$, and thus accounts for
atomic-scale effects. The space dependence of the pair amplitude $F(z)$
was obtained self-consistently by using an efficient numerical algorithm.
From the calculated eigenstates, we were then able to obtain the 
experimentally relevant 
local magnetic moment,
and the local density of states.

We have demonstrated  
that for all the cases considered, where the thickness
$d_S$ of the superconducting layers is
much greater than $\xi_0$ and that of the ferromagnetic
regions is relatively small,
two local minima of the ground state energy
exist, thus yielding
self-consistent states of the 
$0$ and $\pi$ types. Through a careful analysis
of the pair amplitude and excitation spectrum, we have calculated which 
of these two states
is the actual ground state, with
the lowest energy. The results show that the difference
in condensation energies between the $0$ and $\pi$ states 
exhibits damped oscillations
as a function of ferromagnet width, with the characteristic 
exchange-field dependent spatial period 
being given
approximately as $2 \pi (k_{\uparrow}-k_{\downarrow})^{-1}$, the
same quantity which characterizes the oscillations
of $F(z)$ in bilayers.
The local DOS exhibits strikingly 
different behavior  for two exchange fields that
differed by a factor of two. For $I=0.5$, the subgap DOS shows
a gapless structure, with features that depend strongly on
whether the ferromagnet width corresponds to the $0$ or $\pi$ 
state. The half-metallic case ($I=1.0$) is
on the other hand gapless in the range of $d_F$
considered, and  the  modified excitation spectrum  reveals 
itself through the differing peaks in the DOS.
To illustrate the leakage of magnetism into the superconductor,
the differential DOS between the spin up and spin down states
was presented for a SFS junction. The most prominent spin-splitting
was seen for the $\pi$-junction at energies $\epsilon/\Delta_0 \approx 1$.
We believe that this represents
an experimentally important signature for
the $\pi$ state. We have also calculated the local magnetic moment for
both the three and five layer cases, to give further
insight into magnetic polarization effects. Although we found the results
to be relatively insensitive to a $0$ or $\pi$ state configuration, we were 
able to
extract an effective local value of $I(z)$ in both the F and S layers.

The calculations and method used  in this paper, although sufficiently general 
to include in the future
more complicated  effects (e.g. finite  temperature, other
pairing states, spin-flip scattering, and impurities), were  
taken
within the ballistic limit. This limit is appropriate for ferromagnet layers 
whose width
is less than the mean free path, and this is consistent with our calculations, 
where we have taken
$k_S d_F \leq 20$. The inclusion of interfacial scattering 
would likely have the effect of diminishing the proximity effect, without 
qualitatively
altering the characteristic results.  For bulk impurity
scattering, the effective $\xi_F$ would involve not just $I$ but also the
diffusion length. It is the goal of future work 
to address these topics, and also others,
including  heterostructures comprised of a single superconductor sandwiched 
between two ferromagnets with arbitrary relative magnetization,  F/S 
multilayers
with a greater number of layers, and smaller superconductor widths, where
geometrical and atomic-scale effects are likely to be more prevalent.

\begin{acknowledgments}
We thank
 Igor \v{Z}uti\'{c} for many useful discussions.
\end{acknowledgments}


\begin{thebibliography}{11}
\bibitem{altm} Alternative numerical methods e.g.
using infinite Fredholm determinants have
been proposed (see Ref.~\onlinecite{kos} for a survey),
but the precision problems remain.
\end{thebibliography}

\end{document}